\documentclass{emulateapj}
\usepackage{graphicx}
\usepackage{epstopdf}
\usepackage{longtable}

\newcommand{\cii}{[C\,{\sc ii}]}

\newcommand{\mgii}{Mg\,{\sc ii}}

\newcommand{\lya}{Ly$\alpha$}

\newcommand{\hbeta}{H$\beta$}

\newcommand{\asec}{^{\prime\prime}}

\newcommand{\myemail}{chris.willott@nrc.ca}

\def\jwst{{\it James Webb Space Telescope~}}

\def\hersch{{\it Herschel~}}
\def\co21{CO\,(2-1)}

\defcitealias{Willott:2013}{Wi13}

\shorttitle{Star formation rate and dynamical mass of $10^{8}$ solar mass black hole host galaxies}
\shortauthors{Willott et al.}

\begin{document}


\title{Star formation rate and dynamical mass of $10^{8}$ solar mass black hole host galaxies at redshift 6}


\author{Chris J. Willott}
\affil{NRC Herzberg, 5071 West Saanich Rd, Victoria, BC V9E 2E7, Canada}
\email{\myemail}

\author{Jacqueline Bergeron and Alain Omont}
\affil{UPMC Univ Paris 06 and CNRS, UMR7095, Institut d'Astrophysique de Paris, F-75014, Paris, France}




\begin{abstract}

  We present ALMA observations of two moderate luminosity quasars at
  redshift 6. These quasars from the Canada-France High-z Quasar
  Survey (CFHQS) have black hole masses of $\sim 10^{8} M_\odot$. Both
  quasars are detected in the \cii\ line and dust continuum. Combining
  these data with our previous study of two similar CFHQS quasars we
  investigate the population properties. We show that $z>6$ quasars
  have a significantly lower far-infrared luminosity than
  bolometric-luminosity-matched samples at lower redshift, inferring a
  lower star formation rate, possibly correlated with the lower black
  hole masses at $z=6$. The ratios of \cii\ to far-infrared
  luminosities in the CFHQS quasars are comparable with those of
  starbursts of similar star formation rate in the local universe. We
  determine values of velocity dispersion and dynamical mass for the
  quasar host galaxies based on the \cii\ data. We find that there is
  no significant offset from the relations defined by nearby galaxies
  with similar black hole masses. There is however a marked increase
  in the scatter at $z=6$, beyond the large observational
  uncertainties.
\end{abstract}





\section{Introduction}

Improved astronomical observational facilities have enabled the
discovery and study of many galaxies at an early phase of the
Universe's history. It is now possible to witness the majority of the
stellar and black hole mass growth over cosmic time and identify how
physical conditions at early times differ from now. One of the major
relations to be determined as a function of time is the tight
correlation between black hole mass and galaxy properties observed for
nearby galaxies (see \citeauthor{Kormendy:2013} 2013 for a
review). Observations of this relation at high-redshift are critical
to understanding the cause because most of the growth occurred at early
times.

Attempts to measure black hole and galaxy masses at high-redshift face
a number of problems. Black hole mass measurements cannot be made
directly by resolved kinematics of gas or stars within the black
hole's sphere of influence, nor by reverberation mapping. Instead
black hole masses, $M_{\rm BH}$, of quasars can be measured at any
redshift using the single-epoch virial mass estimator that involves
measuring a low-ionization broad emission line, such as \mgii\ or
\hbeta, and calibrating the location of the emitting gas with low-$z$
reverberation-mapped quasars \citep{Wandel:1999a}. For AGN with
obscured broad lines $M_{\rm BH}$ can only be estimated from
the luminosity making an assumption about the accretion rate relative
to the Eddington limit.

Measuring galaxy properties, such as luminosity or velocity
dispersion, $\sigma$, of distant quasars is hampered by surface brightness
dimming, the bright glare of the quasar and AGN (active galactic
nuclei) emission line-contamination of spectral features. Up to
$z\approx 1$ there has been considerable success in measuring AGN host
galaxy luminosities, morphologies and in some cases velocity
dispersions \citep{Cisternas:2011,Park:2014}. At higher redshifts
($1<z<4$) the galaxy light is more difficult to separate from the
quasar, which, combined with greater mass-to-light corrections, lead
to larger uncertainties \citep{Merloni:2010,Targett:2012}. The results
of these studies are mixed with some evidence in favour of higher
$M_{\rm BH}$ at a given galaxy mass. 

At yet higher redshifts it has proved impossible to measure the galaxy
light of quasars \citep{Mechtley:2012} before launch of the \jwst\ and
instead the main method of determining galaxy mass is kinematics of
cool gas in star-forming regions \citep{Carilli:2013}. Facilities such
as the IRAM Plateau de Bure Interferometer, the Jansky Very Large
Array and the Atacama Large Millimeter Array (ALMA) have sufficient
sensitivity and resolution to resolve the gas in distant quasar hosts
and provide dynamical masses
\citep{Walter:2004,Walter:2009,Wang:2010,Wang:2013}. In particular,
ALMA has the sensitivity to probe $z=6$ quasar hosts with star
formation rates, SFR, in the tens of solar masses per year, rather
than only in the extreme starbursts previously observable
\citep{Willott:2013}.  The studies above focussed on $z\approx 6$
Sloan Digital Sky Survey (SDSS) and UKIRT Infrared Deep Sky Survey
(UKIDSS) quasars with high UV and far-IR luminosities and found that
their black holes are on average 10 times greater than the
corresponding $\sigma$ for local galaxies, roughly consistent with a
continuation of the evolution seen in lower redshift studies.

Although observationally there appears to be an increase in $M_{\rm BH}$
with redshift at a given galaxy mass or $\sigma$, it has long been
understood that there are selection biases that affect how closely the
observations trace the underlying distribution. In particular, the
steepness of the galaxy and dark matter mass functions combined with
large scatter in their correlations with black hole mass mean that a
high black-hole-mass-selected sample of quasars will have a systematic
offset in $\sigma$ towards lower values. This
effect, first identified by \citet{Willott:2005b} and
\citet{Fine:2006} was studied in detail in \citet{Lauer:2007} and
numerous studies thereafter. The magnitude of the effect depends upon
the scatter in the correlation, which has not been conclusively
measured at high-redshift, but appears to increase with redshift
\citep{Schulze:2014}. \citet{Willott:2005b} and \citet{Lauer:2007}
showed that the bias is particularly strong for $M_{\rm BH}>10^9 M_\odot$
quasars such as those in the SDSS at $z \approx 6$ and therefore that
the factor of 10 increase in $M_{\rm BH}$ at a given $\sigma$ first seen
in the quasar SDSS\,J1148+5251 \citep{Walter:2004} could be accounted
for by the bias (see also \citeauthor{Schulze:2014} 2014). In
comparison, there would be little bias for a sample of high-$z$
quasars with black hole masses of $M_{\rm BH}\sim10^8 M_\odot$ \citep{Lauer:2007}.

An alternative to measuring the evolution of the assembled galaxy and
black hole masses is to determine the rate at which mass growth is
occurring. For quasars the bolometric luminosity is a measure of the
black hole mass growth rate. For galaxies, the star formation rate is
proportional to the stellar mass growth. The star formation rate can
be determined by the rest-frame far-infrared dust continuum
luminosity. Additionally, the interstellar \cii\ far-infrared emission
line is well-correlated with star-formation
\citep{De-Looze:2014,Sargsyan:2014} so can also be used as a star
formation proxy.

In \citeauthor{Willott:2013} (2013, hereafter Wi13) we presented Cycle
0 ALMA observations in the \cii\ line and 1.2\,mm continuum for two
$z=6.4$ quasars from the Canada-France-High-z Quasar Survey (CFHQS,
\citeauthor{Willott:2010a} 2010b). These quasars have $M_{\rm BH}\sim10^8
M_\odot$, a factor of 10--30 lower than most SDSS quasars known at
these redshifts. One quasar was detected in line and continuum and the
other remained undetected in these sensitive observations placing an
upper limit on its star formation rate of SFR$<40\,M_\odot\,{\rm
  yr}^{-1}$. 

In this paper we present ALMA observations of two further
CFHQS quasars with similar redshift and black hole mass with the aim
of providing a sample large enough to address the issue of how host
galaxy properties such as SFR, $\sigma$ and dynamical mass depend upon
black hole accretion rate and mass at a time just 1 billion years
after the Big Bang. In particular, these quasars are not subject to
the bias in the $M_{\rm BH} - \sigma$ relation discussed previously because of their moderate black hole
masses. Cosmological parameters of $H_0=67.8~ {\rm km~s^{-1}~Mpc^{-1}}$,
$\Omega_{\mathrm M}=0.307$ and $\Omega_\Lambda=0.693$
\citep{Planck-Collaboration:2014} are assumed throughout.

\section{Observations}

CFHQS\,J005502+014618 (hereafter J0055+0146) and CFHQS\,J222901+145709
(hereafter J2229+1457) were observed with ALMA on the 28, 29 and 30
November 2013 for Cycle 1 project 2012.1.00676.S.  Between 22 and 26
12\,m diameter antennae were used. The typical long baselines were
$\sim 400$\,m providing similar spatial resolution to our Cycle 0
observations. Observations of the science targets were interleaved
with nearby phase calibrators, J0108+0135 and J2232+1143. The
amplitude calibrator was Neptune and the bandpass calibrators
J2258-2758 and J2148+0657. Total on-source integration times were
4610\,s for J0055+0146 and 5490\,s for J2229+1457.

The band 6 (1.3\,mm) receivers were set to cover the frequency range
of the redshifted \cii\ transition ($\nu_{\rm rest}$=1900.5369 GHz)
and sample the dust continuum. There are four $\approx 2$\,GHz
basebands, two pairs of adjacent bands with a $11$\,GHz gap in
between. The channel width is 15.625\,MHz (17\,km\,s$^{-1}$).

The data were initially processed by North American ALMA Regional
Center staff with the {\small CASA} software package\footnotemark. On
inspection of these data it became clear that the \cii\ line of
J0055+0146 was located right at the edge of the baseband,
1000\,km\,s$^{-1}$ from the targeted frequency defined by the broad,
low-ionization \mgii\ emission line ($z_{\rm MgII} = 5.983$;
\citeauthor{Willott:2010} 2010a). The \mgii\ line redshift is usually
close to the systemic redshift as measured by narrow optical lines
with a dispersion of 270\,km\,s$^{-1}$ \citep{Richards:2002}. A large
offset for this quasar was not particularly surprising for two
reasons: firstly the signal-to-noise (SNR) of the \mgii\ detection is
not very high and the line appears double-peaked due to noise and/or
associated absorption; secondly the \lya\ redshift ($z_{\rm Ly \alpha}
= 6.02$) is offset from \mgii\ by 1600\,km\,s$^{-1}$ (in the same
direction as the \cii\ offset) and this would make the size of the
\lya\ ionized near-zone negative, which is not physically sensible for
a quasar with such a high ionizing flux and has not been observed in a
sample of 27 $z\approx 6$ quasars \citep{Carilli:2010}. Due to this
redshift uncertainty the receiver basebands were set up so that the
adjacent band covered the \lya\ redshift with zero gap between the two
bands.

\footnotetext{http://casa.nrao.edu}

The default ALMA Regional Center reduction excluded 11 channels at
each end of the 128 channel band. However, only the first 4 channels
need to be excluded, so we re-reduced the data with {\small CASA} to
include more spectral channels at the baseband edges. We checked that
the noise does not increase in these extra channels, except for the
very first and last channels to contain data so we excluded those. In
summary our reduced product contains 118 of the original 128 channels per
baseband, compared with 106 channels in the default reduction.

\begin{figure}[t]
\hspace{-0.5cm}
\includegraphics[angle=0,scale=0.39]{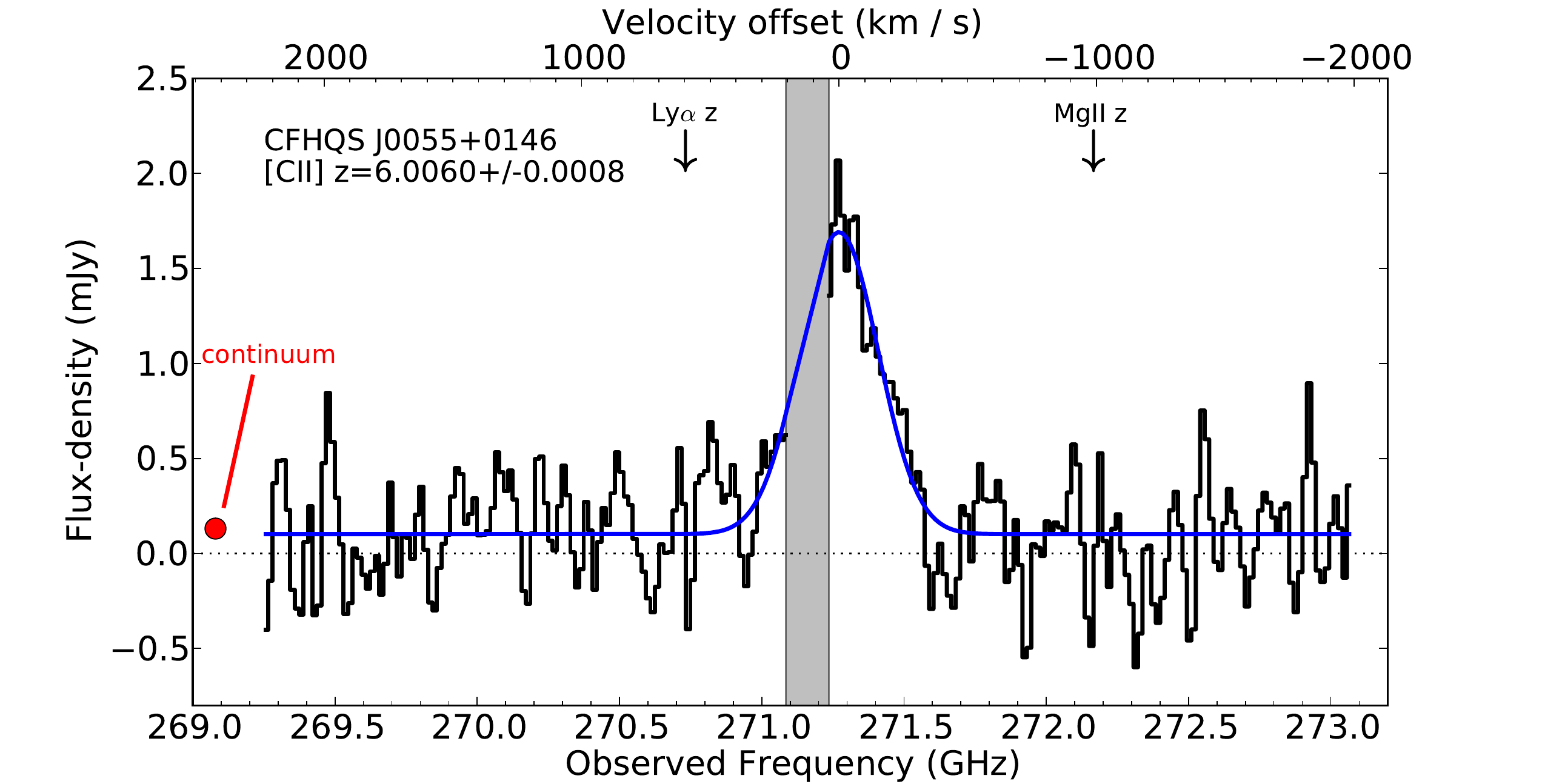}
\caption{ALMA spectrum of J0055+0146 covering two adjacent basebands. The gap between the bands with no data is shaded in gray. The higher frequency band is centred on the redshift determined from the \mgii\ emission line whereas the lower frequency band covers the \lya\ redshift. The \cii\ line is found at the edge of the higher frequency band. The blue curve is a Gaussian plus continuum fit as described in the text. The red circle marks the continuum level independently measured in the three line-free basebands. The upper axis is the velocity offset from the best-fit \cii\ Gaussian peak.}
\label{fig:linespecj0055}
\end{figure}

\section{Results}

Figure \ref{fig:linespecj0055} shows the reduced spectrum of
J0055+0146 from the two adjacent basebands. The final gap between the
bands is only $\approx$ 150 \,km\,s$^{-1}$ and crucially the peak of
the \cii\ line is contained within the higher frequency band. The
lower frequency band contains only a small amount of the line flux but
provides an important constraint on the wings and hence the peak and
width for a symmetric line. A single Gaussian plus flat continuum
model was fit to the available data using a Markov-Chain Monte Carlo
(MCMC). This process shows a good fit for a single Gaussian with
FWHM\,=\,$359 \pm 27$ km\,s$^{-1}$. The formal uncertainty in the FWHM
is very small considering that there is some missing data. This is
because a symmetric line model is used and with the peak and wings
covered by data there is little margin for deviation in the missing
channels.  We add in quadrature an extra 10\% uncertainty in both the
line flux and FWHM due to the missing channels.

\begin{figure}[t]
\hspace{-0.5cm}
\includegraphics[angle=0,scale=0.39]{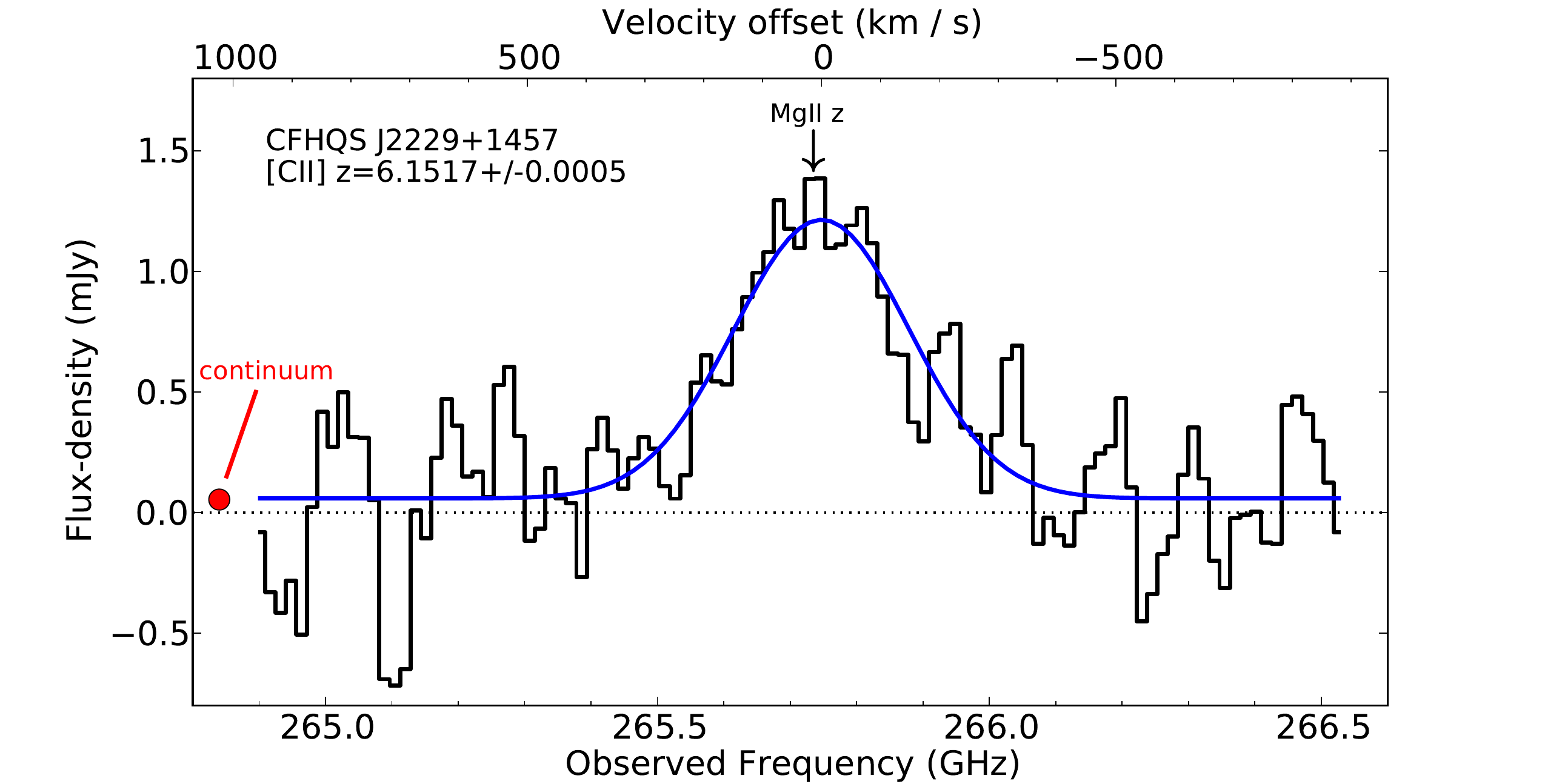}
\caption{ALMA spectrum of J2229+1457 covering the single baseband containing the \cii\ line. The blue curve is a Gaussian plus continuum fit and the red circle the independent continuum level.  The upper axis is the velocity offset from the best-fit \cii\ Gaussian peak.}
\label{fig:linespecj2229}
\end{figure}

The spectrum of J2229+1457 is plotted in Figure
\ref{fig:linespecj2229}. For this quasar the \cii\ line is centred in
the band with no significant offset from the \mgii\ redshift. The line
is consistent with a single Gaussian with a best fit FWHM\,=\,$351 \pm
39$ km\,s$^{-1}$, similar to the value for J0055+0146.  The continuum
level of the fit is again consistent with the independent continuum
level determined from the three line-free basebands. Measurements from
the spectra are given in Table \ref{tab:data}.

\begin{figure}[t]
\includegraphics[angle=0,scale=0.31]{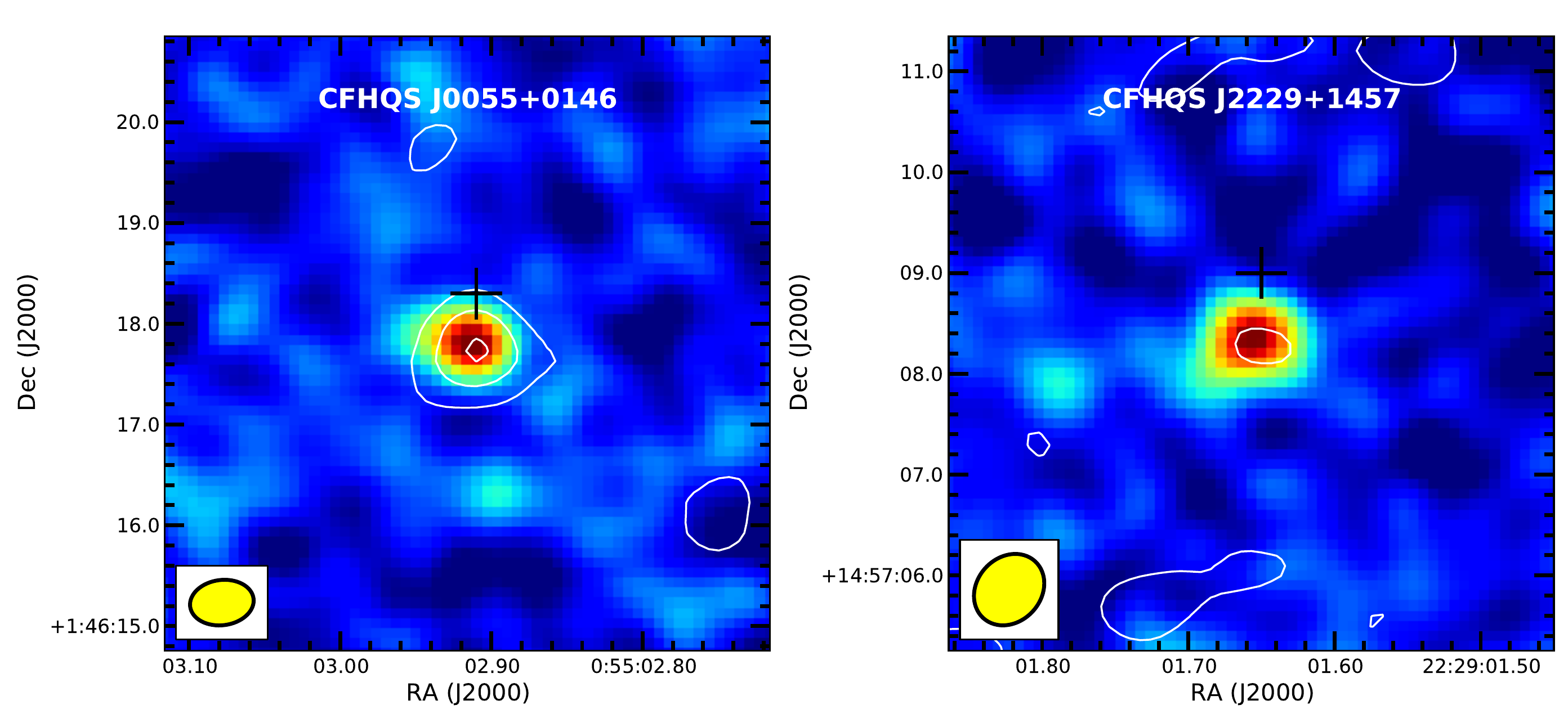}
\caption{The color scale shows the integrated \cii\ line maps for the two quasars. White contours of 1.2\,mm continuum emission from the three line-free basebands are over-plotted at levels of 2,\,4,\,6\,$\sigma$\,beam$^{-1}$. The quasar optical positions are shown with a black plus symbol. The positional offsets between the optical and millimeter are most likely due to astrometric mismatch, rather than a physical offset. J0055+0146 is well-detected in both continuum and line emission. J2229+1457 has only a $2\sigma$ continuum detection that is spatially co-incident with the line emission. The restoring beam is shown in yellow in the lower-left corner.}
\label{fig:contlinemaps}
\end{figure}

Figure \ref{fig:contlinemaps} shows maps of the 1.2\,mm continuum
(white contours) plus the \cii\ line (color scale) for the
quasars. For both quasars there is no significant offset between any
of the continuum or line centroids or the optical quasar position. The
$< 1\asec$ mm-optical offset is within the relative uncertainty of
the optical astrometry. The more accurate \cii\ positions for the two
quasars are 00:55:02.92 +01.46.17.80 and 22:29:01.66 +14.57.08.30. For
J2229+1457 there is only a marginal $2\sigma$ detection of the
continuum located coincident with the peak of the line emission. As
seen in Figure \ref{fig:contlinemaps} there are several other
continuum peaks of this magnitude or greater in the vicinity, so it is
not considered a secure detection, however the measured flux and
uncertainty are included in Table \ref{tab:data}. 

The quasar 1.2\,mm continuum flux-densities were converted to
far-infrared luminosity, $L_{\rm FIR}$, assuming a typical SED for
high-redshift star-forming galaxies. As in \citetalias{Willott:2013}
we adopt a greybody spectrum with dust temperature, $T_{\rm d} =47$\,K
and emissivity index, $\beta=1.6$. To convert from far-IR luminosity
to star formation rate we use the relation SFR $(M_\odot\,{\rm
  yr}^{-1})=1.5\times10^{-10}L_{\rm FIR}\, (L_\odot)$ appropriate for
a Chabrier IMF \citep{Carilli:2013}. We note that this assumes that
all the dust contributing to the 1.2\,mm continuum is heated by hot
stars and not by the quasar.  An alternative estimate of the star
formation rate comes from the \cii\ luminosity. We adopt the relation
in \citet{Sargsyan:2014} of SFR $(M_\odot\,{\rm yr}^{-1}) =
1.0\times10^{-7}L_{\rm [CII]} \, (L_\odot)$.  For the remainder of
this paper, uncertainties on $L_{\rm FIR}$ (and inferred SFR) only
include the flux measurement uncertainties, not that of the dust
temperature and luminosity to SFR conversion.

\begin{table}
\begin{center}
\caption{Millimeter data for the two CFHQS quasars\label{tab:data}}
\vspace{-0.5cm}
\begin{tabular}{lll}
\tableline
& CFHQS\,J0055+0146 & CFHQS\,J2229+1457\\
\tableline
$z_{\rm MgII}\,^{\rm a}$ &  $5.983 \pm 0.004 $ &  $6.152 \pm 0.003 $ \\
$z_{\rm [CII]}$ &  $6.0060 \pm 0.0008$ & $6.1517 \pm 0.0005$\\
FWHM$_{\rm [CII]}$ & $359 \pm 45$ km\,s$^{-1} $& $351 \pm 39$ km\,s$^{-1} $\\
$I _{\rm [CII]} ~($Jy\,km\,s$^{-1}) $ & $0.839  \pm 0.132$ & $0.582 \pm 0.075$\\
$L_{\rm [CII]} ~(L_\odot)$ & $(8.27 \pm 1.30) \times 10^8$  & $(5.96  \pm 0.77) \times 10^8$\\
$f_{\rm 1.2mm}\ (\mu$Jy) & $211 \pm 34$ & $54 \pm 29$\\
$L_{\rm FIR} ~ (L_\odot)$ & $(4.85 \pm 0.78) \times 10^{11}$  & $(1.24 \pm 0.67) \times 10^{11} $\\
SFR$_{\rm [CII]}\,(M_\odot\,{\rm yr}^{-1})$  & $83 \pm 13$  & $60 \pm 8$\\
SFR$_{\rm FIR}\,(M_\odot\,{\rm yr}^{-1})$  & $73 \pm 12$  & $19 \pm 10$\\
$L_{\rm [CII]} / L_{\rm FIR}$ & $(1.70 \pm 0.38) \times 10^{-3}$  &  $(4.80 \pm 2.67) \times 10^{-3}$  \\ 
\tableline
\end{tabular}
\end{center}
{\sc Notes.}---\\
$^{\rm a}$ Derived from \mgii\ $\lambda2799$ observations \citep{Willott:2010}.\\
Uncertainties in $L_{\rm FIR}$,  SFR$_{\rm [CII]}$ and SFR$_{\rm FIR}$ only include measurement uncertainties, not the uncertainties in extrapolating from a monochromatic to integrated luminosity or that of the luminosity-SFR calibrations.
\end{table}

The synthesized beam sizes are $0\farcs63$ by $0\farcs45$ for
J0055+0146 and $0\farcs76$ by $0\farcs64$ for J2229+1457. The better
resolution for J0055+0146 is mostly due to higher elevation of
observation. We used the {\small CASA IMFIT} task to fit 2D gaussian
models to these maps. For J0055 both the continuum and line are
resolved with deconvolved source sizes of $0\farcs51 \pm 0\farcs13$ by
$0\farcs35 \pm 0\farcs26$ at position angle 87 degrees and $0\farcs50
\pm 0\farcs14$ by $0\farcs18 \pm 0\farcs27$ at position angle 62
degrees, respectively. At the distance to this quasar the spatial
extent of $0.5\asec$ is equal to a linear size of 2.9\,kpc. We note
that the missing data in the red wing of the \cii\ line may cause a
bias in the size and inclination if the emission comes from a rotating
disk, but there is no evidence for this based on the similarity of the
line and continuum sizes. For J2229+1457 the continuum is too poorly
detected to attempt a size measurement and the line emission is only
marginally more extended than the beam size. In several other
$z\approx 6$ quasars velocity gradients across the sources are
observed (\citetalias{Willott:2013}; \citeauthor{Wang:2013}
2013). Velocity gradients are not seen for either of these two
quasars, although for J0055+0146 the missing data for
150\,km\,s$^{-1}$ of the red wing hampers our ability to detect such a
gradient.

\section{Discussion}

\subsection{Evolution of far-IR luminosity}
 
In \citetalias{Willott:2013} we reported on the low far-IR
luminosities of the two previously observed CFHQS quasars and
implications for the relatively low SFR of these quasar host galaxies
relative to the black hole accretion rate. We now revisit this issue
with the sample of four $z\approx 6$ CFHQS quasars with ALMA
observations. We note that this sample includes four of the six
  CFHQS quasars with measured black hole masses within the absolute
  magnitude range $-25.5<M_{1450}<-24$ at a declination low enough
for ALMA observation. The two unobserved quasars have
$7\times10^8<M_{\rm BH}<10^9 M_\odot$, and were not observed due to
limited time available and the desire to study the lowest mass black
holes from CFHQS. Therefore there is a slight bias to low black hole
mass in this sample compared to pure UV-luminosity-selection.

Two of the four quasars are well detected in the continuum with fluxes
of $211 \pm 34$ (J0055+0146) and $120 \pm 35$ (J0210-0456)
$\mu$Jy. J2229 has a marginal $2\sigma$ detection of $54 \pm
29\,\mu$Jy (Figure \ref{fig:contlinemaps} and Table \ref{tab:data})
and J2329-0301 is undetected with a $1\sigma$ rms of 30\,$\mu$Jy. We
combine the four values of far-infrared luminosity derived from these
measurements assuming that J2329-0301 has a flux equal to its
$2\sigma$ upper limit of $60\,\mu$Jy. The mean and standard deviation
of the sample is $L_{\rm FIR} = (2.6 \pm 1.4) \times
10^{11}\,L_\odot$. We note this is much lower than the values of
$10^{12} - 10^{13} \,L_\odot$ typically discussed for $z\approx 6$
quasars due to two factors, firstly that the CFHQS sample here have
lower AGN luminosity than most known $z\approx 6$ quasars and a
correlation between AGN and far-IR luminosities is present
\citep{Wang:2011,Omont:2013}, but also that our small sample is
selected on quasar rest-frame UV luminosity and black hole mass,
whereas previous studies have focussed on quasars with pre-ALMA
millimeter continuum detections.

The implication is that these quasars have very high black hole
accretion rates as inferred from the AGN bolometric luminosity,
$L_{\rm Bol}$, yet relatively low SFR. Such a scenario is consistent
with the well-known evolutionary model whereby the optical quasar
phase comes after the main star forming phase
\citep{Khandai:2012,Lapi:2014}, possibly due to quasar feedback
inhibiting gas cooling and star formation. The measured ratio of
$L_{\rm FIR} / L_{\rm Bol}=0.035$ for the four CFHQS quasars is only
found in the optical quasar phase of co-evolution at a time $\sim
1$\,Gyr after the onset of activity for the $z=2$ model of
\citet{Lapi:2014}. Given that this is the age of the universe at $z=6$
the evolution must occur more rapidly at higher redshift. However, the
effect of AGN variability may also be important leading to a selection
effect whereby AGN luminosity-selected objects are observed to have
lower ratios of $L_{\rm FIR} / L_{\rm Bol}$ than the time-averaged
values \citep{Hickox:2014,Veale:2014}.

We have previously shown \citepalias{Willott:2013} that the two
$z=6.4$ CFHQS quasars observed with ALMA in Cycle 0 have $L_{\rm FIR}$
lower than quasars of similar AGN luminosity at lower redshift. On the
other hand, at fixed AGN luminosity $L_{\rm FIR}$ is observed to rise
from $z=0$ to $z=3$
\citep{Serjeant:2010,Bonfield:2011,Rosario:2012,Rosario:2013}.  We
next analyze the evolution of $L_{\rm FIR}$ using our expanded ALMA
sample at $z=6$ and comparable low-redshift data. At all redshifts we
determine the mean $L_{\rm FIR}$ for optically-selected quasars and
X-ray AGN in a narrow range of $L_{\rm Bol}$ corresponding to the mean
$L_{\rm Bol}$ of the four CFHQS $z \approx 6$ quasars in this paper
($L_{\rm Bol} \sim 7 \times 10^{12}\,L_\odot$).

\begin{figure}[t]
\includegraphics[angle=0,scale=0.4]{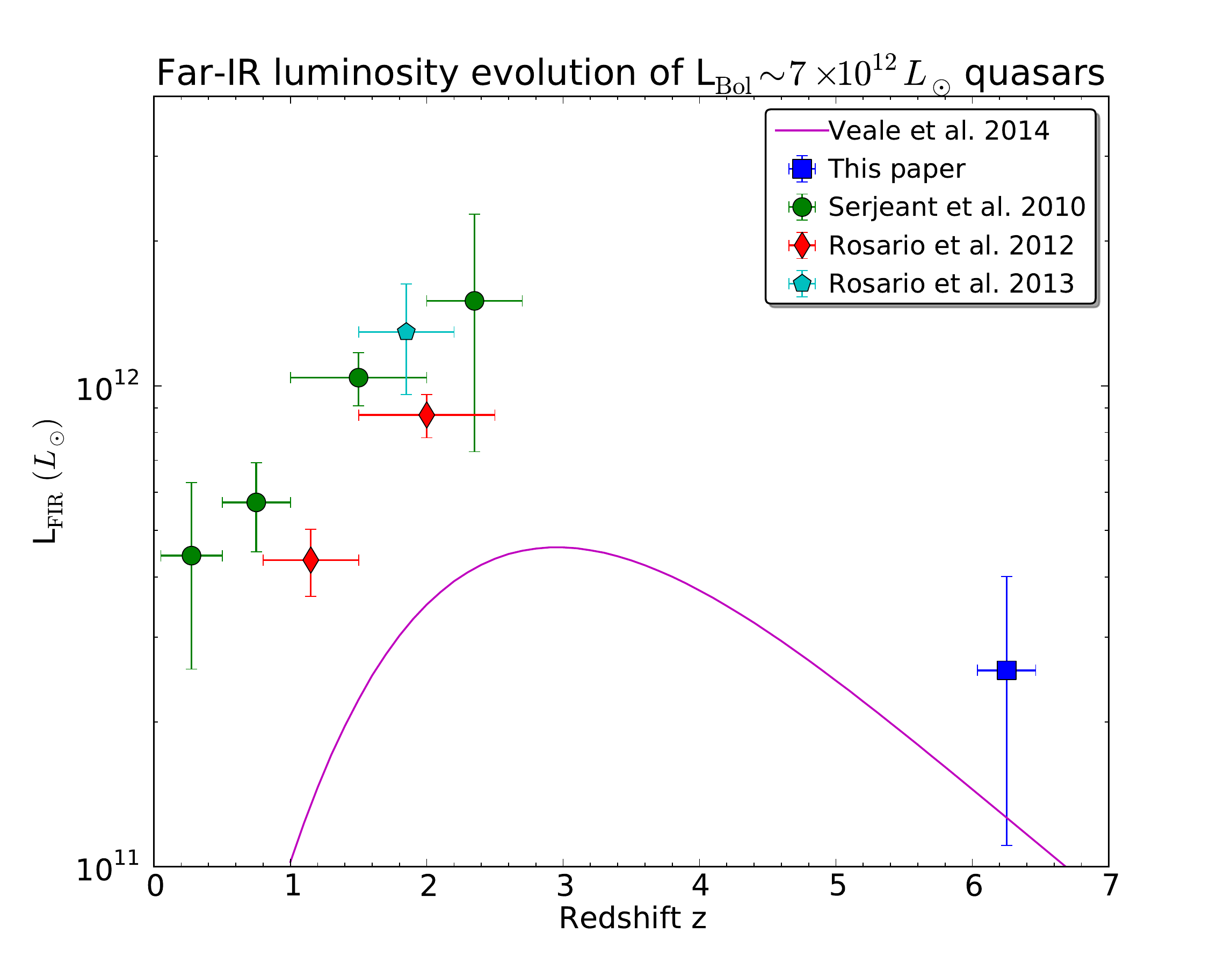}
\caption{Stacked mean far-infrared luminosity for samples of quasars at different redshifts. Details of the samples are described in the text, but all samples are selected to include roughly the same range in bolometric luminosity centred on $L_{\rm Bol} \sim 7 \times 10^{12}\,L_\odot$, the mean $L_{\rm Bol}$ of the four $z\approx 6$ CFHQS quasars plotted with the blue square. There is a clear rise in $L_{\rm FIR}$ up to a peak at $2<z<3$ followed by a decline to $z=6$. The magenta curve shows the mean $L_{\rm FIR}$ due to star formation for the model of \citet{Veale:2014} including scaling by a factor of 2 to account for stellar mass loss.}
\label{fig:lfirevol}
\end{figure}

At $z<3$ we use three datasets based on \hersch imaging of
AGN. \citet{Serjeant:2010} stacked \hersch SPIRE data of
optically-selected quasars and quoted their results as rest-frame
100\,$\mu$m luminosity. We adopt $L_{\rm FIR}= 1.43\,\nu L_{\nu}
(100\,\mu{\rm m})$ \citep{Chary:2001} to convert to far-infrared
luminosity. The absolute magnitude bin $-26<I_{\rm AB}<-24$
corresponds well to $L_{\rm Bol} \sim 7 \times 10^{12}\,L_\odot$
and we have trimmed the size of the highest redshift bin from $2<z<4$
to $2<z<2.7$ because inspection of the luminosity-redshift plane
figure in \citet{Serjeant:2010} shows all but one of the 52 quasars in
this bin are at $z<2.7$. \citet{Rosario:2013} analyzed the \hersch PACS data for optically-selected quasars in COSMOS. Due to
the shorter wavelength of PACS than SPIRE they presented results in
rest-frame 60\,$\mu$m luminosity. We adopt $L_{\rm FIR}= 1.5\,\nu
L_{\nu} (60\,\mu{\rm m})$ \citep{Chary:2001} to convert to
far-infrared luminosity. From this study we use only the highest
redshift, highest luminosity bin as this compares well with $L_{\rm
  Bol} \sim 7 \times 10^{12}\,L_\odot$. \citet{Rosario:2012}
determined the mean infrared luminosity with PACS for X-ray-selected
AGN from the COSMOS survey. We note that the X-ray selected AGN sample
contains a mixture of broad-line, narrow-line and lineless AGN and
these may have different evolutionary properties, but
\citet{Rosario:2013} showed that the mean $L_{\rm FIR}$ of quasars and
X-ray-selected are similar at a given AGN luminosity and redshift.
We use the \citet{Rosario:2012} data from the AGN luminosity bin $8 \times 10^{11}< L_{\rm Bol}< 2
\times 10^{13}\,L_\odot$. Whilst most sources in this bin
have $L_{\rm Bol}< 3 \times 10^{12}\,L_\odot$, we consider the
results appropriate to compare to the $z\approx 6$ quasars as the
correlation between $L_{\rm FIR}$ and $L_{\rm Bol}$ is very shallow at
this luminosity in \citet{Rosario:2012}.

Figure \ref{fig:lfirevol} plots data from these three low-redshift
studies with the CFHQS ALMA bin at $6<z<6.5$. The three low-redshift
studies show a rise in $L_{\rm FIR}$ of a factor of 4 from $z=0.3$ to
$z=2.4$. This rise is attributed to the general increase in massive
galaxy specific star formation rate over this redshift range
\citep{Hickox:2014}. The $6<z<6.5$ bin has a large dispersion due to
the range in 1.2\,mm continuum flux measured for the 4 quasars. The
mean $L_{\rm FIR}$ at $z \approx 6$ is a factor of about 2 lower than
the $z=0.3$ bin and 6 lower than the $z=2.4$ peak at the so-called
{\it quasar epoch}. There is clear evidence here for a turnaround that
mimics the evolution of the quasar luminosity function
\citep{McGreer:2013} and star formation rate density
\citep{Bouwens:2014}, albeit with a much less steep high-redshift
decline due to the fact we are measuring star formation in special
locations within the universe where dark matter halos must have
collapsed much earlier than typical in order to build up the observed
black hole masses of $\sim 10^8\,M_\odot$.

What is the physical reason for this turnaround at $z>3$? In the
evolutionary picture where the optical quasar phase follows the
starburst phase one would expect the star formation and black hole
accretion to be more tightly coupled at high-redshift where there is
barely enough time for star formation to have decreased
substantially. A clue may come from one of the few differences
between quasars at these two epochs. \citet{Willott:2010} showed that
the Eddington ratios of matched quasar luminosity samples at $z=2$ and
$z=6$ are significantly different with the $z=6$ quasars having a
factor of 3$\times$ higher Eddington ratios and therefore 3$\times$ lower black hole
masses than at $z=2$. Such a difference exists between the typical
black hole mass of our CFHQS ALMA sample and that of the highest
luminosity bin of \citet{Rosario:2013}. This Eddington ratio evolution
is observed in other studies
\citep{De-Rosa:2011,Trakhtenbrot:2011,Shen:2012} and predicted by many
theoretical works due to the increase in gas supply to black holes at
high-redshift \citep{Sijacki:2014}.

In Figure \ref{fig:lfirevol} we also plot a theoretical curve of mean
$L_{\rm FIR}$ versus redshift for a simulated sample of rest-frame
UV-selected quasars in the same $L_{\rm Bol}$ range as the observed
quasar samples for the model of \citet{Veale:2014}. This model assumes
an evolving linear relationship between star formation and black hole
growth. The variant of the model plotted here is the ``accretion''
model where the quasar luminosity is proportional to the black hole
growth rate and the Eddington ratio distribution is a truncated
power-law with slope $\beta=0.6$ (dashed curve in Figure 8 of
\citeauthor{Veale:2014} 2014). The model is constrained by the
observed evolving quasar luminosity function and the local ratio of
black hole to galaxy mass. We have scaled this model with a factor of
$2\times$ increase in $L_{\rm FIR}$ to account for stellar mass loss.

As seen in Figure \ref{fig:lfirevol} this curve increases from low
redshift to the peak quasar epoch at $2<z<3$ by about the same factor
as the data, although the total normalization of the curve is lower by
a factor of 3 to 4. \citet{Veale:2014} discuss some of the reasons why
the normalization may be lower than the observations. The decrease in
$L_{\rm FIR}$ with increasing cosmic time from $z=2$ to $z=0$ for
fixed luminosity quasars is due to the fact that such quasars are
rarer at lower redshift and on the steep end of the luminosity
function where scatter is more important. This behaviour also follows
from the general decrease in specific star formation rate with cosmic
time. The high-redshift behaviour of a decline from $z=3$ to $z=6$
matches our observations, so it is instructive to understand why this
occurs in the model. It is due to the assumed $(1+z)^2$ evolution of
the ratio of accretion growth to stellar mass growth, but this assumed
evolution is also degenerate with evolution in the Eddington ratio. As
discussed previously there is observational evidence for positive
evolution in the Eddington ratio from $z=2$ to $z=6$, meaning that the
ratio of accretion growth to stellar mass growth may change more
gradually than $(1+z)^2$ .

A possible alternative explanation for the low $L_{\rm FIR}$ at $z=6$
is that at these early epochs insufficient dust has been generated so
that star formation occurs more often within lower dust environments
\citep{Ouchi:2013,Tan:2013,Fisher:2014,Ota:2014}. In this case there
could be a much smaller decline in the typical SFR of a luminous
quasar hosting galaxy. However, two lines of evidence point towards
this not being the main factor for our quasar sample. First, quasars
at $z=6$ are known to have emission line ratios similar to lower
redshift quasars inferring high metallicity at least close to the
accreting black hole \citep{Freudling:2003}. Second, the \cii\
luminosities in three of the four quasars are high (see below),
suggesting high carbon abundances throughout the host galaxies.

\subsection{The \cii\ -- far-IR luminosity relation}

\begin{figure}[t]
\includegraphics[angle=0,scale=0.4]{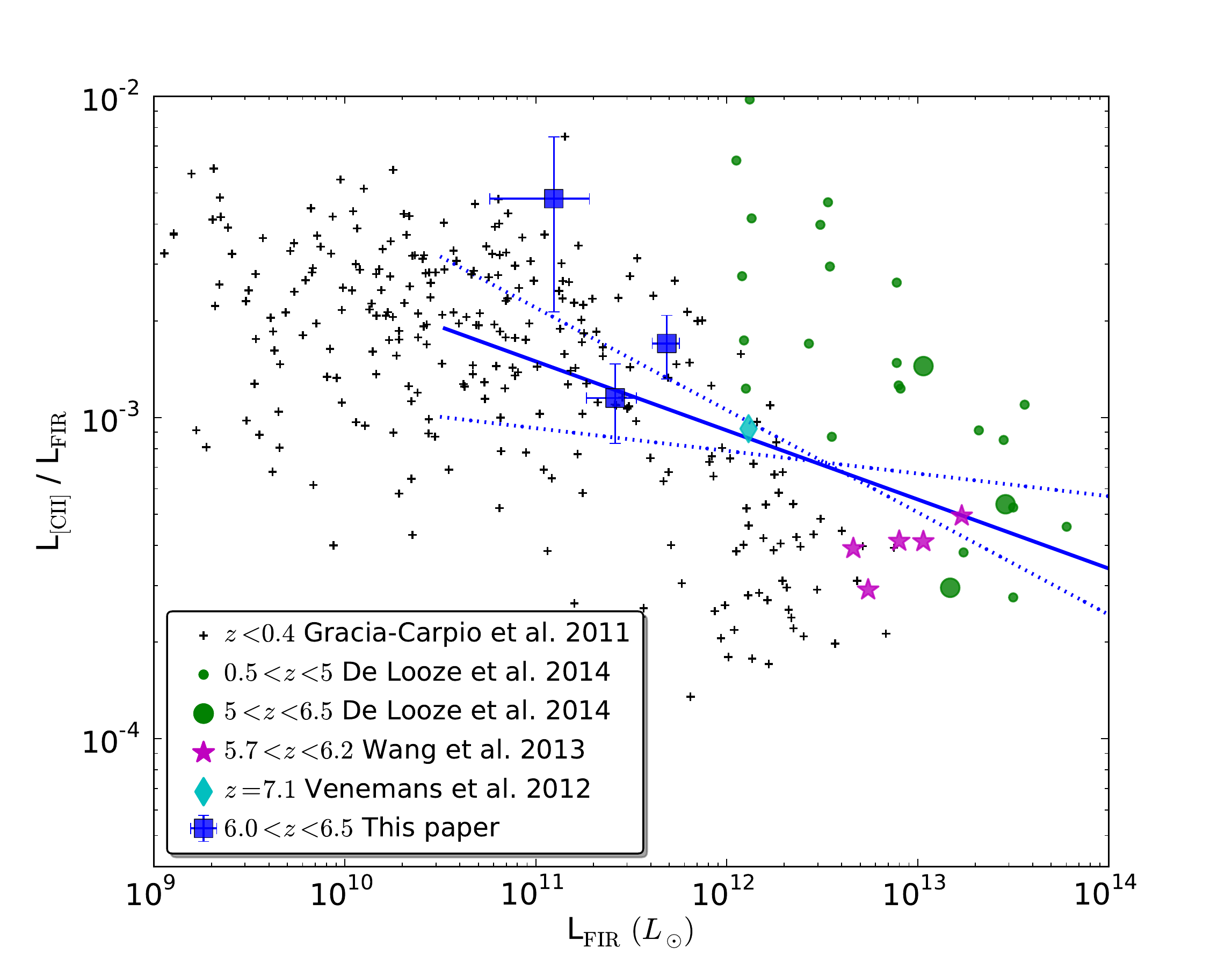}
\caption{The ratio of \cii\ to far-IR luminosity as a function of far-IR luminosity. High-redshift ($z>5$) sources, mostly quasar host galaxies, are identified with large symbols. Error bars are only plotted for the CFHQS ALMA sources to enhance the clarity of the figure. The solid and dotted blue lines show the best fit power-law and $1\sigma$ uncertainty for the  $z>5$ sources. The $z>5$ relation is largely consistent with the distribution of data at lower redshift. } 
\label{fig:lciilfir}
\end{figure}

Three of the four CFHQS ALMA quasars are detected in both \cii\ line
and 1.2\,mm continuum emission (two new detections in this paper and
J0210$-$0456 in \citetalias{Willott:2013}). With the low $L_{\rm FIR}$
discussed in the previous section, these quasar host galaxies probe a
new regime in $L_{\rm FIR}$ at high-redshift. In Figure
\ref{fig:lciilfir} we plot the ratio of \cii\ to far-IR luminosity as
a function of $L_{\rm FIR}$. Also plotted are several samples from the
literature which, due to ALMA at high-redshift and \hersch at
low-redshift, are rapidly increasing in size and data quality. The
low-redshift $z<0.4$ sample of galaxies is from
\citeauthor{Gracia-Carpio:2011} (2011 and in prep.) and contains a mix
of normal galaxies, starbursts and ultra-luminous infrared galaxies
(ULIRGs), some of which contain AGN. The ULIRGs show a {\it \cii\
  deficit} that has been widely discussed in the literature as due to
possible factors including AGN contamination of $L_{\rm FIR}$
\citep{Sargsyan:2012}, high gas fractions \citep{Gracia-Carpio:2011}
or the dustiness, temperature and/or density of star forming regions
\citep{Farrah:2013,Magdis:2014}. Previous observations of high $L_{\rm
  FIR}$ $z>5$ SDSS quasars \citep{Maiolino:2005,Wang:2013} showed a
similar deficit. However many $0.5<z<5$ ULIRGs do not show this
deficit and have $L_{\rm [CII]} / L_{\rm FIR}$ ratios comparable to
low-redshift star-forming galaxies \citep{Stacey:2010}. This is visible in
Figure \ref{fig:lciilfir} for the $0.5<z<5$ compilation of
\citet{De-Looze:2014}.

By adding three $z>6$ quasars with $10^{11}<L_{\rm FIR}<10^{12}\,{\rm
  L}_\odot$ to Figure \ref{fig:lciilfir} we have greatly expanded the
range of luminosities at the highest redshift. Large symbols on Figure
\ref{fig:lciilfir} identify $z>5$ sources. The three $z>5$
\citet{De-Looze:2014} sources are HLSJ091828.6+514223 at $z=5.24$
\citep{Rawle:2014} , HFLS3 at $z=6.34$ \citep{Riechers:2013} and
SDSS\,J1148+5251 at $z=6.42$ \citep{Maiolino:2005}. We note that these
sources were mostly selected for followup based on high $L_{\rm FIR}
$. The quasar ULAS\,J1120+0641 at $z=7.1$ has a more moderate $L_{\rm
  FIR}$ and $L_{\rm [CII]} / L_{\rm FIR}$ ratio and lies in between
the CFHQS and high $L_{\rm FIR}$ objects on the plot.

Although we are wary of the selection effects in Figure
\ref{fig:lciilfir} and the as yet unknown cause for the change in
$L_{\rm [CII]} / L_{\rm FIR}$ with $L_{\rm FIR}$ at high-redshift, the
new data provide the opportunity to make the first measurement of the
slope of this relation at $z>5$. We fit the 12 $z>5$ sources with a
single power-law model of the dependence of $L_{\rm FIR}$ on $L_{\rm
  [CII]}$ incorporating the observational (but not systematic)
uncertainties using a MCMC procedure. The best-fit
relation is
\begin{equation}
\log_{10} L_{\rm FIR}=0.59+1.27 \log_{10} L_{\rm [CII]}.
\end{equation}
Therefore the $L_{\rm [CII]} / L_{\rm FIR}$ ratio line plotted in
Figure \ref{fig:lciilfir} has a logarithmic slope of
$(1/1.27)-1=-0.21$. The MCMC $1\sigma$ uncertainties, based solely on
the observational data, are plotted as dotted lines. These also favor
a shallow negative slope, not nearly as steep as the slope that would
be fit to the previous $z>0.5$ data that covers only a narrow range of
$L_{\rm FIR}$. The $L_{\rm [CII]} / L_{\rm FIR}$ ratios of the CFHQS
and ULAS $z>6$ quasars are not greatly different to those of similar
$L_{\rm FIR}$ galaxies at low-redshift. 

\citet{Wang:2013} note that the low $L_{\rm [CII]} / L_{\rm FIR}$
ratios for SDSS $z>5$ quasars may be at least in part due to AGN
contamination of the far-IR emission. Future observations at higher
spatial resolution will be critical to examine differences in the
spatial distribution of the line and continuum emission
(e.g. \citeauthor{Cicone:2014} 2014). We expect to observe that the
dust continuum is more compact than the \cii\ line in the high $L_{\rm
  FIR}$ $z>5$ quasars, due to either more centrally-concentrated
starbursts with higher dust temperatures, like local ULIRGs, or AGN
dust-heating. In contrast we expect the low $L_{\rm FIR}$ $z>5$
quasars have star formation spread more evenly throughout their host
galaxies, with similar spatial distribution of line and continuum emission.

\subsection{The  $z\approx 6$ $M_{\rm BH}-\sigma$ and $M_{\rm BH}-M_{\rm dyn}$  relationships}

The combination of black hole mass estimates and \cii\ line host
galaxy dynamics for these $z\approx 6$ quasars allows us to
investigate the black hole - galaxy mass correlation at an early epoch
in the universe. The evolution of this relationship is a critical
constraint on the co-evolution (or not) of galaxies and their nuclear
black holes. As discussed in the Introduction there are reasons to
believe that past studies using only the most massive black holes from
SDSS quasars (e.g. \citeauthor{Wang:2010} 2010) were prone to a bias
where one would expect the black holes to be relatively more massive
than the galaxies \citep{Willott:2005b,Lauer:2007}, as observed. With
new data on $M_{\rm BH} \sim 10^8\,M_\odot$ quasars we are able to
test this hypothesis and determine any real offset from the local
relationship. Additionally, most previous work in this area has used
the molecular CO line to trace the gas dynamics. CO is usually more
centrally concentrated than \cii, so \cii\ potentially probes a larger
fraction of the total mass (although we note that
\citeauthor{Wang:2013} 2013 found similar dynamical masses using CO
and \cii\ for their $z\approx 6$ quasars).

In addition to the CFHQS and \cite{Wang:2013} quasars we add to our
study two other $z>6$ quasars observed in the \cii\ line:
SDSS\,J1148+5251 at $z=6.42$ and the most distant known quasar,
ULAS\,J1120+0641 at $z=7.08$. Both these quasars have \mgii-derived
black hole masses \citep{De-Rosa:2011,De-Rosa:2014} with very low
measurement uncertainties. The black hole masses for all three CFHQS
quasars also come from \mgii\ measurements \citep{Willott:2010}. Some
of these spectra are of moderate SNR and have substantial measurement
uncertainties on the black hole masses. To all the quasars with
\mgii-derived black hole masses we add a 0.3 dex uncertainty to the
measurement uncertainties to account for the dispersion in the
reverberation-mapped quasar calibration \citep{Shen:2008}. 

None of the \cite{Wang:2013} quasars have \mgii\ measurements so black
hole masses are estimated assuming that the quasars radiate at the
Eddington limit, as observed for most $z\approx 6$ quasars
\citep{Jiang:2007,Kurk:2007,Willott:2010,De-Rosa:2011}.  The
dispersion in the lognormal Eddington ratio distribution at $z \approx
6$ is 0.3 dex \citep{Willott:2010}. We add 0.3 dex uncertainty from
the observed dispersion in the Eddington ratio distribution in
quadrature to the 0.3 dex due to the dispersion in the
reverberation-mapped quasar calibration for a total uncertainty on the
\cite{Wang:2013} quasar black hole masses of 0.45 dex.

First we consider the $M_{\rm BH}-\sigma$ relationship. For nearby
galaxies $\sigma$ is the velocity dispersion of the galaxy bulge. At
high-redshift bulges are less common \citep{Cassata:2011} and we do
not expect the $z\approx 6$ kinematics to match that of a pressure
supported bulge. With the limited spatial resolution of current data
we cannot be sure the \cii\ gas is distributed in a rotating disk,
although there is evidence of this for some sources \citep{Wang:2013}.
\cite{Ho:2007a} discusses the relationship and calibration of bulge
velocity dispersion and disk circular velocity and concludes that
although there is additional scatter one can relate molecular or
atomic gas in a disk to stellar bulges. The major complication is the
inclination of the disk. For a random sample of inclinations this can
be modelled, however there is a possibility that quasars have disks
oriented more often face-on, reducing the line-of-sight velocity
dispersion \citep{Ho:2007}. 

\begin{table*}
\begin{center}
\caption{Mass determinations for $z>5.7$ quasars\label{tab:masses}}
\vspace{-0.2cm}
\begin{tabular}{lccccl}
\tableline
Name                         & $z_{\rm [CII]}$                 & $M_{\rm BH} ~(M_\odot)\,^{\rm a}$ & $\sigma ^{\rm b}$ & $M_{\rm dyn}~(M_\odot)$ & Refs.$^{\rm c}$\\
\tableline
 CFHQS\,J0055+0146 & $6.0060 \pm 0.0008$ & $(2.4^{+2.6}_{-1.4})\times 10^{8} $  & $207 \pm 45$ & $4.2\times 10^{10} $  & 1,2 \\
CFHQS\,J0210$-$0456 & $6.4323 \pm 0.0005$  &  $(0.8^{+1.0}_{-0.6})\times 10^{8} $ & $98 \pm 20$ & $1.3\times 10^{10} $  & 1,2,3 \\
CFHQS\,J2229+1457 & $6.1517 \pm 0.0005$  & $(1.2^{+1.4}_{-0.8})\times 10^{8}$  & $241 \pm 51$ & $4.4\times 10^{10} $ & 1,2 \\
SDSS\,J0129$-$0035   & $5.7787 \pm 0.0001$  & $(1.7^{+3.1}_{-1.1})\times 10^{8} $ & $112 \pm 21$ & $1.3\times 10^{10} $  & 4 \\
SDSS\,J1044$-$0125   & $5.7847 \pm 0.0007$  & $(1.1^{+1.9}_{-0.7})\times 10^{10} $ & $291 \pm 76$ &  --- $^{\rm d}$ &  4 \\
SDSS\,J1148+5251   & $6.4189 \pm 0.0006$  & $(4.9^{+4.9}_{-2.5})\times 10^{9} $ & $186 \pm 38$ &  $1.8\times 10^{10} $ & 5,6,7\\
SDSS\,J2054$-$0005   & $6.0391 \pm 0.0001$ & $(0.9^{+1.6}_{-0.6})\times 10^{9} $ & $364 \pm 67$ & $7.2\times 10^{10} $  &  4 \\
SDSS\,J2310+1855  & $6.0031 \pm 0.0002$ &$(2.8^{+5.1}_{-1.8})\times 10^{9} $  & $325 \pm 61$ & $9.6 \times 10^{10} $  & 4 \\
ULAS\,J1120+0641  &  $7.0842 \pm 0.0004$ & $(2.4^{+2.4}_{-1.2})\times 10^{9} $ & $ 144 \pm 34$ &   $2.4\times 10^{10} $ & 8,9\\
ULAS\,J1319+0950  & $6.1330 \pm 0.0007$   &$(2.1^{+3.8}_{-1.4})\times 10^{9} $  & $381 \pm 91$ &  $12.5\times 10^{10} $ &  4 \\
\tableline
\tableline
\end{tabular}
\end{center}
{\sc Notes.}---\\
$^{\rm a}$ Derived from \mgii\ $\lambda2799$ observations if possible, else from Eddington luminosity assumption. Uncertainties include observational errors plus systematics based on calibrations.\\
$^{\rm b}$ Derived from Gaussian FWHM fit to \cii\ spectrum using method of \cite{Ho:2007} including an inclination correction (see text for individual inclinations assumed). Uncertainties include observational errors plus systematics based on calibrations.\\
$^{\rm c}$ References: (1) This paper, (2) \citet{Willott:2010}, (3) \citet{Willott:2013}, (4) \citet{Wang:2013}, (5) \citet{Maiolino:2005}, (6) \citet{Walter:2009}, (7) \citet{De-Rosa:2011}, (8) \citet{Venemans:2012}, (9) \citet{De-Rosa:2014}.\\
$^{\rm d}$ This quasar does not have a dynamical mass calculation in \citet{Wang:2013} due to the difference in the \cii\ and CO line profiles.
\end{table*}

We determine $\sigma$ using the method of \cite{Ho:2007}, specifically
setting the \cii\ line full-width at 20\% equal to 1.5$\times$ the
FWHM as expected for a Gaussian since most of the lines are
approximately Gaussian. The \cii\ emitting gas is assumed to be in an
inclined disk where the inclination angle, $i$, is determined by the
ratio of minor ($a_{\rm min}$) and major ($a_{\rm maj}$) axes,
$i=\cos^{-1}( a_{\rm min}/a_{\rm maj})$. The circular velocity is
therefore $v_{\rm cir}=0.75 \,$FWHM$_{\rm [CII]} / \sin i$. For all of
the quasars in this study we determine an inclination from the \cii\
data or assume an inclination if one or both the major and minor axes
are unresolved. All the quasars in \cite{Wang:2013} were spatially
resolved, although some had quite large uncertainties on $a_{\rm min}$
and $a_{\rm maj}$. We adopt the inclination angles from their paper.

For the \cii\ emission of J0210$-$0456, $a_{\rm maj} =0\farcs52 \pm
0\farcs25$ (2.9\,kpc) with $i=64^{\circ}$
\citepalias{Willott:2013}. For J0055+0146, $a_{\rm maj} =0\farcs50 \pm
0\farcs14$ (2.9\,kpc) with $i=69^{\circ}$ (Section 3, assuming no bias
from missing red wing data). The \cii\ emission of J2229+1457 is only
marginally spatially resolved ($0\farcs85$ versus beam size of
$0\farcs76$) and we estimate an intrinsic FWHM of $\approx 0\farcs4$
(2.4\,kpc). An inclination of $i=55^{\circ}$ is assumed as this is the
median inclination angle for the resolved sources in this paper and
\cite{Wang:2013}. Neither SDSS\,J1148+5251 nor ULAS\,J1120+0641 have
published inclination angles, so we also assume $i=55^{\circ}$ for
both of them. We adopt FWHM$_{\rm [CII]} =287 \pm 28$\,km\,s$^{-1}$
for SDSS\,J1148+5251 \citep{Walter:2009} and FWHM$_{\rm [CII]} =235
\pm 35$\,km\,s$^{-1}$ for ULAS\,J1120+0641 \citep{Venemans:2012}
. Values of black hole masses and $\sigma$ for this sample are
provided in Table \ref{tab:masses}.

\begin{figure}[t]
\includegraphics[angle=0,scale=0.4]{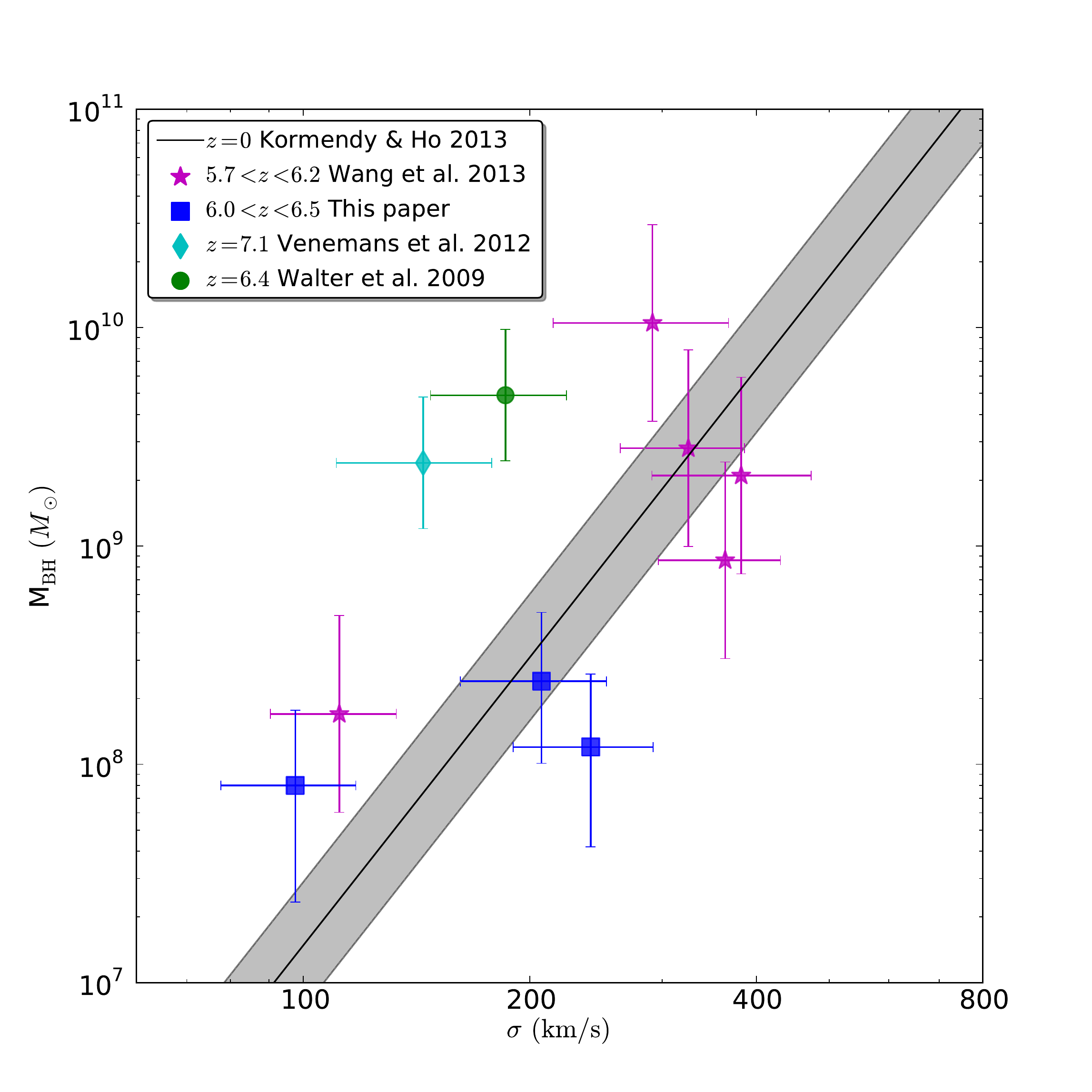}
\caption{Black hole mass versus velocity dispersion calculated from the \cii\ line using the method of \cite{Ho:2007} for $z\approx 6$ quasars. Quasars from the CFHQS are shown as blue squares and the other symbols show quasars from the SDSS and ULAS surveys. The black line with gray shading is the local correlation $\pm 1\sigma$ scatter of black hole mass and bulge velocity dispersion \citep{Kormendy:2013}. The $z\approx 6$ quasars are distributed around the the local relationship, but with a much larger scatter and some quasars with significantly lower $\sigma$ for their $M_{\rm BH}$.} 
\label{fig:mbhsig}
\end{figure}

Figure \ref{fig:mbhsig} shows the $M_{\rm BH} - \sigma$ relationship
for the $z\approx 6$ quasar sample. Uncertainties on black hole masses
include the scatter in the calibration as described
previously. Uncertainties in $\sigma$ include FWHM measurement
uncertainty plus a 10\% uncertainty for the conversion from FWHM to
$v_{\rm cir}$ and 15\% for the conversion from $v_{\rm cir}$ to
$\sigma$ as seen in the sample of \cite{Ho:2007a}. The black line is the
local correlation of \cite{Kormendy:2013} with the gray band the $\pm
1\sigma$ scatter. The first thing to note is that the quasars are
distributed around the local relationship rather than all being offset
to low $\sigma$ as is commonly believed to be the case. As noted by
\cite{Wang:2010}, using the method of \cite{Ho:2007}, rather than
calculating $\sigma$ as FWHM$/2.35$, leads to much higher
$\sigma$. Note that this is without adopting extreme face-on
inclinations for most quasars. There are still several quasars, such
as SDSS\,J1148+5251 and ULAS\,J1120+0641, that have values of $\sigma$
considerably lower than the local relation.

The main result of Figure \ref{fig:mbhsig} is that whilst there is
little mean shift between the $z=0$ and $z\approx 6$ data, there is a
much larger scatter in the data at $z\approx 6$, well beyond the size
of the error bars. This larger scatter at an early epoch is expected
based on dynamical evolution, incoherence in AGN/starburst activity
and the tightening of the relation over time from merging
\citep{Peng:2007}. We note that our hypothesis that the bias described
in the Introduction would lead to the lower $M_{\rm BH}$ quasars being
located on the local relation with a lower scatter than the high
$M_{\rm BH}$ quasars is not supported by these observations. The
scatter in $\log_{10} \sigma$ at $M_{\rm BH}\approx 10^8\,M_\odot$ is
about the same as that at $M_{\rm BH}> 10^9\,M_\odot$

We go one step further from $\sigma$ to determine dynamical masses
using the deconvolved \cii\ sizes. For consistency, we follow the
method of \cite{Wang:2013}.  The dynamical mass within the disk radius
is given by $M_{\rm dyn} \approx 1.16\times 10^5\, v_{\rm cir}^2 \,D
\,M_\odot$ where $D$ is the disk diameter in kpc and calculated as
$1.5 \times$ the deconvolved Gaussian spatial FWHM.  The resulting
dynamical masses are given in Table \ref{tab:masses}. We note that
there is considerable uncertainty on these values due to the unknown
spatial and velocity structure of the gas, the marginal spatial
resolution and limited sensitivity that means we may be missing more
extended gas. Due to the these uncertainties we do not place formal
error bars on the dynamical masses, following \cite{Wang:2013}. Higher
resolution data in the future are required to confirm the derived
masses.

\begin{figure}[t]
\includegraphics[angle=0,scale=0.4]{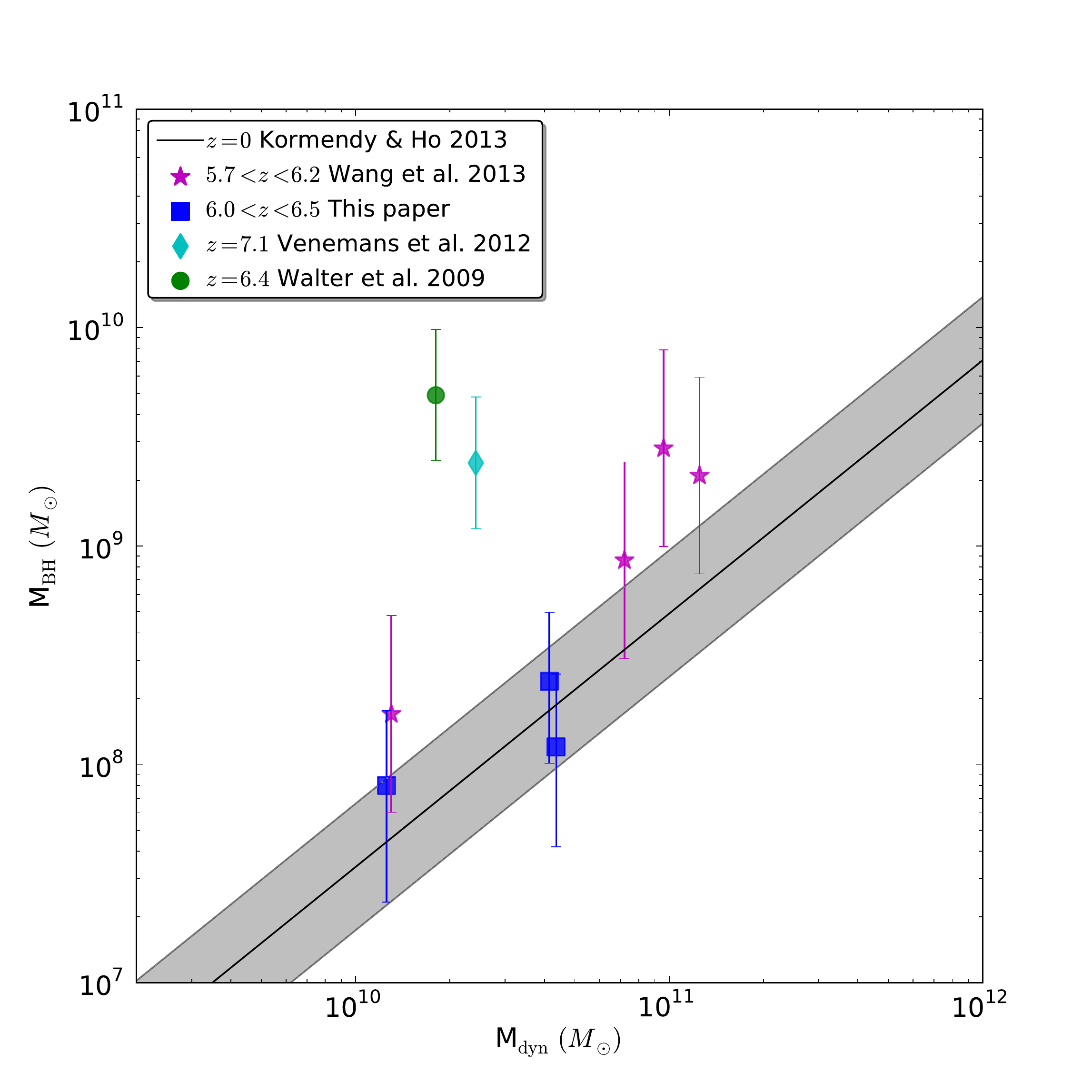}
\caption{Black hole mass versus host galaxy dynamical mass for $z\approx 6$ quasars. Symbols as for Figure \ref{fig:mbhsig}. The black line with gray shading is the local correlation $\pm 1\sigma$ scatter from the work of \cite{Kormendy:2013} equating $M_{\rm dyn}$ to $M_{\rm bulge}$. The CFHQS quasars lie on the local relationship and do not show the large offset displayed by the most massive black holes. Uncertainties in $M_{\rm dyn}$ have not been calculated due to the reasons given in the text.} 
\label{fig:mbhmdyn}
\end{figure}

SDSS\,J1148+5251 has been extensively studied in
\cii\ \citep{Maiolino:2005,Walter:2009,Maiolino:2012,Cicone:2014}. The
highest resolution observations by \cite{Walter:2009} revealed a very
compact circumnuclear starburst with radius 0.75\,kpc and FWHM$_{\rm
  [CII]} =287 \pm 28$\,km\,s$^{-1}$. For an assumed inclination of
$i=55^{\circ}$ this gives $M_{\rm dyn}=1.8 \times
10^{10}\,M_\odot$. For comparison \cite{Walter:2004} determined a
dynamical mass from CO emission in this quasar of $5.5\times
10^{10}\,M_\odot$ within a larger radius of 2.5\,kpc, the larger
radius being the main difference between the results. Recent
observations have shown more complex \cii\ emission including evidence
for gas extended over tens of kpc and at high velocities indicative of
outflow \citep{Maiolino:2012,Cicone:2014}. We adopt $M_{\rm dyn}=1.8
\times 10^{10}\,M_\odot$ for SDSS\,J1148+5251 noting that the true
value could be several times larger. The most distant known quasar,
ULAS\,J1120+0641 at $z=7.08$, has been well detected in \cii, although
not yet spatially resolved \citep{Venemans:2012}. Based on the
published FWHM$_{\rm [CII]} =235 \pm 35$\,km\,s$^{-1}$ and assuming a
spatial FWHM of 3\,kpc (similar to the other quasars resolved by ALMA)
and $i=55^{\circ}$ we determine $M_{\rm dyn}=2.4 \times
10^{10}\,M_\odot$.

In Figure \ref{fig:mbhmdyn} we plot black hole mass versus galaxy
dynamical mass for the most distant known quasars. The black line and
gray shading represent the local correlation of $M_{\rm BH}$ with
bulge mass $M_{\rm bulge}$ \citep{Kormendy:2013}. In the absence of gas
accretion and mergers the present stellar bulge mass represents the
sum of the gas and stellar mass at high-redshift, so it is a good
comparison for the dynamical mass within the central few
kpc. \cite{Kormendy:2013} note that their correlation (their equation
10) gives a black hole to bulge mass ratio of 0.5\% at $M_{\rm
  bulge}=10^{11}\,M_\odot$ that is 2 to 4 times higher than previous
estimates due to the omission of pseudobulges, galaxies with uncertain
$M_{\rm BH}$ and ongoing mergers.

The position of the high-$z$ data with respect to low redshift is
fairly similar to Figure \ref{fig:mbhsig}, not surprising because
$v_{\rm cir}$ derived from the \cii\ velocity FWHM is a major factor
in both $\sigma$ and $M_{\rm dyn}$. The points are shifted somewhat
further from the local bulge mass than for the local velocity
dispersion. This shift is due to the smaller size of galaxies at
high-redshift, as the size is the only term in the derivation of
dynamical mass not in $\sigma$. We note the much greater dynamic range
in black hole mass (2 dex) than in dynamical mass (1 dex) in our
sample. This is likely due more to our selection over a wide range of
quasar luminosity than to a non-linear relationship between these
quantities at $z=6$.

All three of the CFHQS quasars lie within the local $1\sigma$ scatter
and the one $M_{\rm BH}\sim 10^8\,M_\odot$ quasar in \cite{Wang:2013}
is only a factor of 4 greater than the local relationship. In contrast
the $M_{\rm BH}> 10^9\,M_\odot$ quasars tend to show a larger scatter
and larger offset above the local relationship as previously found
\citep{Walter:2004,Wang:2010,Venemans:2012,Wang:2013}.  We caution
that there are considerable uncertainties in some of these
measurements as already discussed, but in dynamical mass the results
look more like we would expect based on the quasar selection bias
effect.

\section{Conclusions}

During ALMA Early Science cycles 0 and 1 we have observed a complete
sample of four $z>6$ moderate luminosity CFHQS quasars with black hole
masses $\sim 10^8\,M_\odot$. Three of the four are detected in both
far-IR continuum and the \cii\ emission line. The far-IR luminosity is
found to be substantially lower than that of similar luminosity
quasars at $1<z<3$. Assuming that far-IR luminosity traces star
formation equally effectively at these redshifts this implies that at
$z\approx 6$ quasars are growing their black holes more rapidly than
their stellar mass compared to at the peak of
the {\it quasar epoch} ($1<z<3$).

The ratios of [CII] to far-IR luminosities for the CFHQS quasars lie
in the range 0.001 to 0.01, similar to that of low-redshift galaxies
at the same far-IR luminosity. This suggests a similar mode of
star-formation spread throughout the host galaxy (rather than in dense
circumnuclear starburst regions that have lower values for this ratio
in local ULIRGs). Combining with previous $z>5.7$ quasar data at
higher $L_{\rm FIR}$ we find that the far-IR luminosity dependence of
the \cii/FIR ratio has a shallow negative slope, possibly due in part
to an increase in $L_{\rm FIR}$ due to quasar-heated dust in some
optically-luminous high-$z$ quasars.

The three CFHQS quasars well-detected in the \cii\ emission line allow
this atomic gas to be used as a tracer of the host galaxy dynamics.
Combining with published data on higher black hole mass quasars we
have investigated the $M_{\rm BH}-\sigma$ and $M_{\rm BH}-M_{\rm dyn}$
relations at $z\approx 6$. We show that the $z=6$ quasars display a
$M_{\rm BH}-\sigma$ relation with similar slope and normalization to
locally, but with much greater scatter. Similar results are obtained for the $M_{\rm
  BH}-M_{\rm dyn}$ relation with a somewhat higher normalization at
$z=6$ and a higher scatter at high $M_{\rm BH}$. As discussed in 

Combining our results on the relatively low $L_{\rm FIR}$ for $ M_{\rm
  BH} \sim 10^8\,M_\odot$ $z\approx 6$ quasars with their location on
the $M_{\rm BH}-\sigma$ relation leads to something of a paradox. The
fact these quasars lie on the local $M_{\rm BH}-\sigma$ relation
suggests that their host galaxies have undergone considerable
evolution to acquire such a high dynamical mass. So why is it that
this mass accumulation is not leading to a high star formation rate?
As discussed in \citetalias{Willott:2013}, simulations such as those
of \citet{Khandai:2012} and \citet{Lapi:2014} predict that such low ratios of SFR
to black hole accretion occur after episodes of strong feedback that
inhibits star formation throughout quasar host galaxies.  Another
possibility mentioned in Section 4.1 is that $L_{\rm FIR}$ fails to
trace star formation so effectively in these high-redshift galaxies,
due to lower dust content (e.g. \citeauthor{Ouchi:2013} 2013). Note
that using $L_{\rm [CII]}$ as a star formation rate tracer instead of $L_{\rm FIR}$,
would give higher SFR by a factor of three for one of the CFHQS
quasars.

Higher resolution follow-up \cii\ observations of these quasars are
critical to measure more accurately the distribution and kinematics of
the gas used as a dynamical tracer in order to reliably determine the
location and scatter of the correlations between black holes and their
host galaxies at high-redshift.


\acknowledgments

Thanks to staff at the North America ALMA Regional Center for
processing the ALMA data. Thanks to Melanie Veale for useful
discussion and providing her models in electronic form. This paper
makes use of the following ALMA data:
ADS/JAO.ALMA\#2012.1.00676.S. ALMA is a partnership of ESO
(representing its member states), NSF (USA) and NINS (Japan), together
with NRC (Canada) and NSC and ASIAA (Taiwan), in cooperation with the
Republic of Chile. The Joint ALMA Observatory is operated by ESO,
AUI/NRAO and NAOJ. The National Radio Astronomy Observatory is a
facility of the National Science Foundation operated under cooperative
agreement by Associated Universities, Inc.



{\it Facility:} \facility{ALMA}.



\bibliography{willott}

\end{document}